\documentstyle[11pt,newpasp,twoside]{article}

\def\edcomment#1{\iffalse\marginpar{\raggedright\sl#1\/}\else\relax\fi}
\reversemarginpar

\begin{document}
\title{Optical and infrared interferometry}
\author{Timothy R. Bedding}
\affil{School of Physics, University of Sydney 2006, Australia}

\begin{abstract}
Interferometric techniques are at the forefront of modern astronomical
instrumentation.  A new generation of instruments are either operating or
nearing completion, including arrays of small telescopes as well as the
``big guns'' (VLTI and Keck).  A number of space interferometers for the
detection of extra-solar planets are also being planned.  I will review the
current state of play and describe the latest developments in the field.
\end{abstract}

\section{Introduction}

It is somewhat disappointing that a meeting on ``Galaxies and their
Constituents at the Highest Angular Resolutions'' contains so few results
from optical inter\-ferometry\footnote{In this review, the term ``optical''
is also meant to include infrared wavelengths.}.  To some extent, this
reflects the interests of those on the Scientific Organising Committee, but
it must also be taken as a sign that optical interferometry has yet to
deliver on its promises.  Why is this?  One reason is surely the
difficultly of the technology and its vulnerability to Hofstadter's Law
(1980):
\begin{quotation}
{\em Hofstadter's Law} --- It always takes longer than you expect, even when
you take into account Hofstadter's Law.
\end{quotation}

But progress is being made, and there is a large number of interferometry
projects in various stages of development and operation.  The following is
the current list of operating interferometers, listed in the order in which
they obtained first fringes:
\begin{itemize}  \itemsep=0pt
\item     GI2T/REGAIN (Grand Interf\'erom\`etre \`a 2 T\'elescopes).
\item     SUSI (Sydney University Stellar Interferometer)
\item     COAST (Cambridge Optical Aperture Synthesis Telescope)
\item     ISI (Infrared Spatial Interferometer) 
\item     FLUOR (Fiber Linked Unit for Optical Recombination) 
\item     IOTA (Infrared-Optical Telescope Array)
\item     NPOI (Navy Prototype Optical Interferometer)
\item     PTI (Palomar Testbed Interferometer)
\item     CHARA  (Center for High Angular Resolution Astronomy)
\end{itemize}
The following are under construction (in alphabetical order, since I am not
game to predict which will be first to get fringes!):
\begin{itemize}  \itemsep=0pt
\item Keck Interferometer

\item LBT (Large Binocular Telescope)

\item MIRA-II (Mitaka IR Array)

\item VLTI (Very Large Telescope Interferometer)

\end{itemize}
Finally, there are planned space missions, including SIM, TPF and
DARWIN\@.  

More details of all these projects can be found via links from the
Web-based ``Optical Long Baseline Interferometry News,'' currently
maintained by Peter Lawson\footnote{\tt http://huey.jpl.nasa.gov/olbin/}.
Here, I simply want to stress the number and range of the projects, and to
point out that many of them are producing scientific results.  The
proceedings of recent conferences (Unwin \& Stachnik 1999; Lena \&
Quirrenbach 2000) testify to the vigorous activity in this field.  In
Section~\ref{sec.recent}, I highlight some areas in which recent
technological developments have been made.  First, however, it is
appropriate to make some general remarks about optical interferometry.

\section{Interferometers measure fringes!}

Realising that interferometers measure fringes is the key to understanding
their capabilities and limitations (see, e.g., Bedding 1997;
Quirrenbach 1997).  To observe an object, an interferometer must
be able to detect fringes and track them.  This requires that the object
has a sufficiently compact bright component, or that a suitable reference
source (e.g., a star) is located nearby on the sky.

The main observable is the fringe contrast -- the visibility -- which is a
dimensionless number between zero and one that indicates the extent to
which a source is resolved on the baseline being used.  Importantly, the
signal-to-noise with which the visibility can be measured is a function of
$NV^2$, where $N$ is the number of photons detected per sub-aperture per
integration time, and $V$ is the {\em measured\/} visibility (including the
inevitable reduction by atmospheric and instrumental effects).  This
dependence on $NV^2$ implies that interferometry becomes increasingly
difficult for faint sources, but also for those with complex structures.
For example, a drop in visibility from 1 to 0.1 as a source becomes
resolved is equivalent, in terms of signal-to-noise, to a drop in
brightness by a factor of 100 (5 magnitudes)!

The $NV^2$ limit can be addressed in several ways:
\begin{itemize} \itemsep=0pt

\item Using larger sub-apertures.  However, this requires adaptive optics
if the telescope diameter becomes much larger than~$r_0$, the average scale
over which the wavefront is flat.

\item Bootstrapping to increase integration times, which can be achieved by
tracking fringes over a connected series of short baselines to allow
low-visibility fringes to be measured on the longest.  This approach
requires lots of telescopes, which must be deployed rather wastefully in
terms of $(u,v)$ coverage.

\item Tracking fringes on a point source to increase integration time on
the target.  This is best done with a dual-feed system.

\end{itemize}

\section{Recent developments}	\label{sec.recent}

\subsection{Fibres and integrated optics}

Single-mode optical fibres have been used very successfully for spatially
filtering light beams (Coude Du~Foresto et~al.\ 1997).  This
allows visibility amplitudes to be measured to high precision (about one
percent), by reducing the sensitivity to atmospheric fluctuations.  So far,
results have been demonstrated with IOTA in the $K$ band (2.2 micron;
Perrin et~al.\ 1999) and now in the $L$~band (3.75\,micron;
Mennesson et~al.\ 1999).  It seems clear that fibres have an
important role to play in interferometry, especially in the infrared.

A related development is in the field of integrated optics, which
potentially allows large tables of bulk optics to be replaced by miniature
devices.  White-light interference has been achieved in the laboratory with
two (Berger et~al.\ 1999) and three beams (Haguenauer
et~al.\ 2000).  It is too soon to estimate the impact of this technology
on optical interferometers, with applications in space perhaps being the
most likely.

\subsection{Mid-infrared interferometry}

The only interferometer operating in the mid-infrared is the ISI, which has
a heterodyne system.  A recent development has been the application of ISI
to spectral-line observations of molecules around giant stars
(Monnier et~al.\ 2000).  Both VLTI and Keck are targeting the
mid-infrared, which is very attractive because the 8--10-m apertures are
essentially diffraction-limited at 10--20\,microns.  A lot of excellent
science can be expected from those instruments, provided the problems with
thermal background can be overcome, but the amount of observing time will
obviously be limited.

\subsection{Imaging}

Imaging is often seen as the main goal of interferometry, despite the fact
that a lot of scientific questions can be answered without reconstructing
an image.  We must not expect high-quality VLA-type images from optical
interferometers, given the much smaller number of array elements and the
much larger atmospheric phase fluctuations.  Instead, the correct parallel
to draw is with VLBI, in which closure-phase imaging is a well-established
technique.  The first crude images from optical arrays have been produced
for binary stars (Baldwin et~al.\ 1996; Hummel et~al.\
1998) and barely-resolved single stars (Young et~al.\ 2000).

The key to good imaging is a large number of array elements.  This is well
known by radio astronomers, and the potential in the optical is shown by
the spectacular images produced with aperture-masking of single telescopes
such as the Keck (Tuthill et~al.\ 2000).  Images from current
long-baseline arrays will not be this good, since none has more that six
elements.  The next generation of imaging interferometers will need at
least ten, and preferably fifteen elements.

\subsection{Differential phase measurements}

Apart from fringe visibility, the other main observable for an
interferometer is fringe phase.  With three or more telescopes, closure
phases can be used to reconstruct complex objects, as mentioned above.  But
even with two telescopes, useful phase measurements can be made.  These
involve measuring the differential phase of one object relative to another
(or one object relative to itself at different wavelengths), which allows
very precise astrometry.  A nice feature of differential phase measurements
is that much interesting science can be done with point sources (i.e.,
stars), so that the $NV^2$ limitation is overcome.  Examples include
parallaxes and proper motions, such as small motions induced by the
gravitational pull of unseen planets and, for dual-wavelength observations,
the shift in photo-centre caused by the presence of a ``hot Jupiter''
planet.

The key technological element is a dual feed, in which two stars are
observed simultaneously so that their differential phase can be measured.
Such a system also allows dual-wavelength operation, in which a single star
is observed at two wavelengths with the two feeds.  Finally, the second
feed can instead be used for fringe tracking on a reference star, giving
longer integration times on a fainter and/or more complex source.  All
these techniques have been demonstrated for the first time with the PTI
(Lane et~al.\ 2000) and, although problems remain, it appears that plans to
apply these techniques to the Keck Interferometer and space missions are
feasible.

\subsection{Nulling}

Methods for using an interferometric null to allow a faint object (e.g., a
planet) to be seen next to a bright star are being developed by many groups
(Hinz et~al.\ 1998; Guyon et~al.\ 1999;
B{\"o}ker \& {Allen} 1999; Velusamy et~al.\ 1999;
Ollivier et~al.\ 2000; Wallace et~al.\ 2000).  Deep and
stable nulls have now been demonstrated in the laboratory and with
telescopes, and nulling capabilities will be included in the Keck
Interferometer.  The scientific aims include direct detection of
exoplanets, and also of exo-zodiacal dust.  Eventually, it is hoped that
space interferometers will provide images and even spectra of Earth-like
exoplanets.  And that somewhat optimistic -- but perhaps achievable --
scenario seems a good place to end this review!

\acknowledgments For travel support, I thank the Australian Research
Council and the Science Foundation for Physics in the University of
Sydney.

\end{document}